\documentstyle[12pt]{article}
 \def\avd#1{\overline{#1}}

 \def\s{\sigma}

 \newcommand{\be}{\begin{equation}}
 \newcommand{\ee}{\end{equation}}
 \title{On the concept of complexity in random dynamical systems}
 \author{Vittorio Loreto$^1$, Giovanni Paladin$^2$
 and Angelo Vulpiani$^1$}
 
\begin{document}
 \maketitle
 \centerline{$^1$Dipartimento di Fisica, Universit\`a di Roma
 'La Sapienza'}
 \centerline{P.le A.Moro 2 I-00185 Roma, Italy}
 \centerline{$^2$Dipartimento di Fisica,  Universit\`a dell'Aquila}
 \centerline{Via Vetoio I-67100 Coppito, L'Aquila, Italy}
 \date{}
 \medskip

 \begin{abstract}
  We introduce a measure of complexity
 in terms of the average number of bits per time unit necessary
 to specify the sequence generated by the system.
 In random dynamical system, this indicator
  coincides with the rate $K$ of divergence of nearby trajectories
  evolving under two different noise realizations.
 The meaning of $K$ is discussed in the context of the
  information theory, and it is shown that it can be determined
  from real experimental data.
  In presence of strong dynamical intermittency,
 the value of  $K$ is very different from the standard Lyapunov exponent
 $\lambda_\sigma$ computed considering two nearby
 trajectories evolving under the same randomness.
 However, the former is much more relevant than the latter
 from a physical point of view as illustrated by some numerical
  computations for noisy maps and sandpile models.
 \end{abstract}
 \vskip .4truecm
 \noindent
 PACS NUMBERS: 05.45.+b

 \section{Introduction}

 In deterministic dynamical systems there exist well established ways
 to define the complexity of a temporal evolution in terms of
 Lyapunov exponents and Kolmogorov-Sinai entropy.

 However,  the situation becomes much more ambiguous in presence
 of a random perturbation, or more in general a generic randomness,
 which are always present in physical systems as a consequence
 of thermal fluctuations or uncontrollable changes of
 control parameters and in numerical experiments because of
 the roundoff errors \cite{a1}.

 In the literature, a first rough conclusion is that
 the presence of a small noise does not change the qualitative behaviour of
 the dynamics \cite{a2}.
 In the case of a regular (stable) system, the random
 perturbation just changes the very long time behavior by introducing
 the possibility of jumps among different attractors
 (stable fixed points, stable limit cycles or tori).
 A familiar example is the Langevin equation describing the motion of
 an overdamped particle in a double well.

 Even in the opposite limit of chaotic dissipative systems,
 the presence of noise is expected not to change the qualitative behavior
 in a dramatic way. The typical situation is the following:

 \begin{enumerate}
 \item[a)] the strange attractor maintains the fractal structure
 at larger scales, although it is smoothed at small
 scale $O(\sigma)$, if $\sigma$ is the strength of the noise;

 \item[b)] the value of Lyapunov exponents differs from the
 unperturbed one of a quantity $O(\sigma)$.
 \end{enumerate}

 However the combined effects of the noise and of the deterministic part
 of the evolution law can produce highly non-trivial, and often
 intriguing, behaviours \cite{a3}-\cite{a8}.
 Let us mention the stochastic resonance where there is a synchronization
 of the jumps between two stable points \cite{a9}-\cite{a11} and the
 phenomena of the so called noise-induced order
 \cite{a7} and  of the noise-induced instability \cite{a5}-\cite{a6}.

 In our opinion one of the main problem
 is the lacking of a well define method to
 characterize the ``complexity''  of the trajectories.
 Usually \cite{a2, a5, a7}, the degree of chaoticity is measured
 by treating the random term as a usual time-dependent term,
 and therefore, considering the separation of two nearby trajectories
 with the same realization of the noise. In this way it is possible
 to compute the maximum Lyapunov exponent $\lambda_{\sigma}$
 associated to  the separation rate of two nearby trajectories
 with the same realization of the stochastic term.

 Some authors thus argue that there exists
 a phenomenon of noise-induced order \cite{a7}, when
 at increasing the strength of the fluctuation $\sigma$,
  $\lambda_{\sigma}$ passes from positive to negative.
 Even the opposite phenomenon (noise-induced instability) has
 been observed: at increasing $\sigma$, $\lambda_{\sigma}$ can
 pass from negative to positive \cite{a5}-\cite{a6}.

 Although the Lyapunov exponent $\lambda_{\sigma}$ is a well defined
 quantity, it is neither unique nor the most useful
 characterization of complexity.
 In addition, a moment of reflection shows that it is practically
 impossible to extract $\lambda_{\sigma}$ from experimental data.

 In this paper we introduce a more natural indicator of complexity
 in random dynamical systems  computing the separation rate
 of nearby trajectories evolving in
 two different realizations of the noise, instead of only one.
Let us stress that, such a procedure exactly corresponds
 to what happens when experimental data are analyzed by
 the Wolf et al. algorithm \cite{a12}.
 Basically, our measure of complexity is related to the average number
 per time unit of bits necessary to specify the sequence generated
 by a random evolution law.

 The outline of the paper is the following.

In sect.II we introduce
 the simplest way to treat the randomness by discussing two specific
 examples: the Langevin equation describing the motion of an
 overdamped particle in a double well and the case of the so-called
 stochastic resonance. These two examples provide a clear evidence
 of the limitations that arise when the the
 Lyapunov exponent is computed by treating the noise term as a usual time
 dependent term as well as of the necessity of a better characterization
 of the complexity of "noisy" systems.

 Sect.III and IV are devoted to the definition of an appropriate indicator
 of complexity respectively for dynamical systems with noise and random
 dynamical systems. For this last case, where the randomness is not
 simply given by an additive noise, we discuss two examples of systems
 which can be described by random maps: a two block earthquake model
 \cite{lacorata} and sandpile models \cite{bak2},
 an interesting example of Self-Organized Criticality \cite{bak1}.

 The basic features of random maps are discussed in a
 a one-dimensional map, which exhibits interesting
 behaviours like the so-called {\em on-off intermittency} \cite{onoff}.

 Sect.V  discusses in detail the case of Sandpile models with respect to
 the definition of complexity and to the predictability problem.

 In sect.VI we discuss the results and we draw the conclusions.

 \section{The naive approach: the noise treated as a standard
 function of time}

 The simplest information about the chaoticity of noisy systems
 can be obtained treating the random term as a usual
 time-dependent term, and therefore, considering the
 separation of two nearby trajectories with the same
 realization of the noise. Such a characterization can
 be misleading, as illustrated in the following example.

 \subsection{Langevin Equation}

Let us consider the one-dimensional Langevin equation
 \be
 { {d x} \over {d t} }= - { {\partial V(x)} \over {\partial x}}
                        +   \sqrt{2\sigma}\, \eta
 \ee
 where $V(x)$ diverges for $\mid x\mid \to \infty$ and it has more than
  one minimum, e.g. the usual double well $V=-x^2/2+x^4/4$, and
  $\eta(t)$ is a white noise.

 The Lyapunov exponent $\lambda_{\sigma}$
  associated to  the separation rate of two nearby trajectories
 with the same realization of the stochastic term $\eta(t)$, is

 \be
  \lambda_{\sigma}=\lim_{t\to \infty} { 1 \over t} \ln |z(t)|
 \label{deflyap}
 \ee
 where the evolution of the tangent vector (that should be regarded
 as an infinitesimal perturbation of the trajectory $x(t)$) is:

 \be
 { {d z} \over {d t} }= - { {\partial^2 V(x(t))} \over {\partial x^2}}z(t).
 \ee

 Since the system is ergodic with invariant probability distribution
 $P(x)=C e^{-V(x)/\sigma}$, one has:

 \be
 \begin{array}{l}
  \lambda_{\sigma}=\lim_{t\to \infty} { 1 \over t} \ln |z(t)|=
 -\lim_{t\to \infty} { 1 \over t} \int_0^t
  \partial^2_{xx} V(x(t')) dt'=\\
 -C \int \partial^2_{xx} V(x) e^{-V(x)/\sigma} \,\, dx=
 -{C \over \sigma } \int (\partial_x V(x))^2
  e^{-V(x)/\sigma} \,\, dx < 0
 \end{array}
 \ee

 This result is rather intuitive: the trajectory $x(t)$ spends
 most of the time in one of the "valleys" where
 $-\partial^2_{xx} V(x) < 0$
 and only short periods on the "hills" where
 $-\partial^2_{xx} V(x) > 0$, so that there is a decreasing
 of the average of the logarithm of the distance between two
 trajectories evolving in the same noise realization.
 Let us remark that in Ref. \cite{nicolis}, using a wrong argument,
 an opposite result is claimed.

 As matter of fact, $\lambda_{\sigma}<0$ implies a fully predictable
 process ONLY IF the realization of the noise is known.
 In the case, more sensible, of two
 initially close trajectories evolving in two different
 noise realizations, after a certain time $T_{\sigma}$, the two
 trajectories will be very distant since they will be in two different
 "valleys". For $\sigma \to 0$, by
 the Kramer formula, one has  $T_\sigma \sim \exp  \Delta V/\sigma$
  where $\Delta V$ is the
 difference between the values of $V$ on the top of the hill and
 on the bottom of the valley.

 The result obtained for the one dimensional Langevin equation can
 be easily generalized at any dimension for gradient systems
 supposing that the noise is small enough.

 Let us consider the system

 \be
 \dot x_i = - \frac{\partial V}{\partial x_i} + \sqrt{2 \sigma} \eta_i
 \ee
 where $<\eta_i (t) \eta_j (t^\prime)>= \delta_{i,j} \delta (t-t^\prime)$.
 Denoting with $R^2= \sum \vert z_i \vert^2$ one has:

 \be
 \frac{1}{2}\frac{d R^2 (t)}{dt}= - \sum_{i,j} z_i
 \frac{\partial^2 V}{\partial x_i \partial x_j} z_j=
 - ({\displaystyle \bf z}(t), \hat A(t) {\displaystyle \bf z}(t)) \le -
l({\displaystyle \bf x}(t)) R^2 (t)
 \ee
 where $l({\displaystyle \bf x})$ is the minimum eigenvalue of the matrix
 $\hat A$ whose elements are

 \be
 A_{i,j}= \frac{\partial^2 V}{\partial x_i \partial x_j}.
 \ee

 For the Lyapunov exponent $\lambda_{\sigma}$ one obtains

 \be
 \lambda_{\sigma} = \lim_{t \rightarrow \infty} \frac{1}{2t} ln
 \frac{R^2 (t)}{R^2 (0)}  \le - \lim_{t \rightarrow \infty}
 \frac{1}{t}   \int_0^t l({\displaystyle \bf x}(t^\prime)) dt^\prime =
 - \frac{1}{C} \int l({\displaystyle \bf x}) \exp^{-V/\sigma}
d {\displaystyle \bf x}.
 \ee

 Since $l({\displaystyle \bf x})$ is positive around the minimum it follows
that
 $\lambda_{\sigma} < 0$ for small values of $\sigma$. From a more
 rigorous discussion see \cite{scop}.

 \subsection{Stochastic resonance with and without noise}

 Let us now discuss a deterministic systems close to the onset of
 chaos when the control parameter varies periodically in time.
 We consider the set of differential equations which is a slight
 modification of the Lorenz model \cite{lorenz}

 \be
 \left\{
 \begin{array}{l}
 dx/dt= 10(y-x)\\
 dy/dt= -xz + R(t)x - y\\
 dz/dt= xy -{8 \over 3}z
 \end{array}
 \right.
 \label{season}
 \ee
 where the control parameter has a periodic time variation:

 \be
 R(t)=R_0 -A \, \cos(2 \pi t/T).
 \ee

 In our case, the periodic variations of $R$ roughly  mimic the
 seasonal changing on the solar heat inputs.

 An interesting situation is when the average Rayleigh number
 $R_0$ is assumed to be close to the threshold $R_{cr}=24.74$
 for the transition from stable fixed points to a chaotic attractor
 in the standard Lorenz model. The value of the amplitude $A$ of the
 periodic forcing  should be such  that $R(t)$ oscillates below and
 above $R_{cr}$.
 For very large $T$, a good approximation of the  solution is given by

 \be
 x(t)=y(t)=\pm \sqrt{ {8\over 3} (R(t)-1)} \;\;\;\; z(t)=R(t)-1
 \ee
 which is obtained by the fixed points of the standard Lorenz model
 by replacing $R$ by $R(t)$.
 The stability of this solution is a rather complicated issue,
 which depends on the values of $R_0$, $A$ , and $T$.

 If $R_0$ is larger than $R_{cr}$
 the solution is unstable. In this case, for $A$ large enough
 (at least $R_0-A < R_{cr}$)
 one observes a mechanism similar to that of the stochastic resonance
 in bistable systems with random forcing.
 The value of $T$ is crucial: for large $T$ the systems behaves as
 follows. Let us introduce

 \be
     T_n \simeq nT/2 - T/4
 \ee
 the times at which $R(t)=R_{cr}$.

 For $0<t<T_1$, the control parameter  $R(t)$ is smaller than
 $R_{cr}$ so that the system is stable and the trajectory
 is close to one of the two solutions (eq.\ref{season}).
 For $ T_1<t<T_2$, one has  $R(t)>R_{cr}$
 and both solutions (eq.\ref{season}) are unstable so that the trajectory
 in a short time relaxes toward a sort of `adiabatic' chaotic attractor.
 The chaotic attractor smoothly changes at varying $R$
 above the threshold $R_{cr}$, but if $T$ is large enough,
 this dependence can be neglected in a first approximation.
 However, when $R(t)$ becomes again smaller than $R_{cr}$,
 the `adiabatic' attractor disappears and, in general, the system is
 far from the stable solutions (eq.\ref{season}).
 But, since they are attracting, the system relaxes toward them.
 See figure (1.a)

 For a detailed analysis of this behaviour see ref \cite{crisanti}.

 It is worth stressing that  the system is chaotic, i.e.
 the first Lyapunov exponent is positive, although
 the correlation function of the variable $z$ does not decay
 as a consequence of strong correlation
 between the regular intervals.

 Let us now discuss the effect of
 a random forcing, of strength $\sigma$, in the case where
  $R(t)-R_{cr}$  changes sign during the time evolution but the
 solutions (eq.\ref{season}), in the absence of the noise, are stable.
 In practice, we consider the Langevin equation

 \be
 \left\{
 \begin{array}{l}
 dx/dt= 10 \, (y-x) + \sqrt{2\sigma}\, \eta_1\\
 dy/dt= -x \, z + R(t) \, x - y +\sqrt{2\sigma}\, \eta_2\\
 dz/dt= x \, y -{8\over 3} \, z +\sqrt{2\sigma}\, \eta_3
 \end{array} \right.
 \label{langevin}
 \ee

 where $\eta_i(t)$ are uncorrelated white noises i.e.
 $<\eta_i(t)\eta_j(t')>=\delta_{ij} \delta(t-t')$.

 The numerical study of the model  reveals a
 phenomenology very close to the original stochastic resonance.
  For small values of $\sigma$ one has
 the same qualitative behavior obtained
 at $\sigma=0$, while for $\sigma$ slightly larger than
 a critical value $\sigma_{cr}$ one has an alternation of regular
 and irregular motions.  See figure 1.b.
 Now  the Lyapunov
 exponent, computed treating the noise as an usual time-dependent term,
 is negative, i.e. two trajectories, initially close, with the
 same realization of the random forcing do not separate
 but stick exponentially fast.

 Figures (1.a) and (1.b) show in a clear way how, if noise is
 involved, one can obtains simulations rather close either in the case
 of a positive Lyapunov exponent (fig.1.a)  or whit a negative Lyapunov
exponent (fig.1.b).

 The two above examples show the limitation of the Lyapunov
 exponent computed treating the noise term as a usual time dependent term
 for the characterization of the "complexity" of noisy systems.

 \section{Complexity in dynamical systems with noise}

 The main difficulties to define the notion of complexity
 in a deterministic evolution law with a random perturbation
 already appear in 1D maps.
 In fact, the generalization to $N$-dimensional maps
 or to coupled ordinary differential equations is straightforward.

 Let us therefore consider the model map

 \be
    x(t+1)=f[x(t),t]+\sigma w(t)
 \ee
 where $t$ is an integer and $w(t)$ is an uncorrelated random process,
  e.g. $w$ are independent random variables uniformly
 distributed in $[-1,1]$.
 The maximum Lyapunov exponent $ \lambda_{\sigma}$ defined in (\ref{deflyap})
  is given by the map for the tangent vector:
 \be
   z(t+1)=f'[x(t),t] \, z(t) \
 \ee
 where $f'=df/dx$
 At $\sigma=0$,    $\lambda_0$ is the Lyapunov exponent
  of the unperturbed map.

 In order to introduce a more natural indicator of complexity
  in noisy dynamics it is convenient to follow a quite different approach,
  where two realizations of the noise, instead of only one,
 are used \cite{paladin}.

 Before discussing our alternative definition of chaos in noisy systems,
 we must briefly  recall what are the characterization of intermittency
  in deterministic dynamical systems.
 An effective Lyapunov exponent \cite{a13} has been introduced to
  measure the fluctuations of chaoticity
 \be
 \gamma_t(\tau) = {1\over \tau} \ln {|z(t+\tau)| \over |z(t)|}
 \ee
 It gives the local expansion rate in the interval
 $[t, t+\tau]$. The maximum Lyapunov exponent is thus
 given by a time average along the trajectory $x(t)$:
 $
 \lambda_0= <\gamma_t> \qquad {\rm for} \ \tau \to \infty
 $.

  Let us define the new indicator
  of complexity  in the framework of
  the  deterministic map with no random perturbation
  where it coincides with $\lambda_0$.
 Let $x(t)$ be  the trajectory starting at $x(0)$ and $x'(t)$
  be the trajectory starting at $x'(0)=x(0)+\delta x (0)$
  with $\delta_0=|\delta x(0)|$
  and indicate by $\tau_1$ the maximum time such that
   $|x'(t)-x(t)|<\Delta$. Then, we put
 $x'(\tau_1+1)=x(\tau_1+1)+\delta x(0)$and define  $\tau_2$ as
 the maximum $\tau$ such that $|x'(\tau_1+\tau)-x(\tau_1+\tau)|<\Delta$
 and so on.
  In our context, we can define the effective Lyapunov  as
 \be
 \gamma_i = {1\over \tau_i} \ln {\Delta\over \delta_0}
 \ee
 However, we sample the expansion rate
  in a non-uniform way,  at time intervals $\tau_i$.
  As a consequence the probability of picking
 $\gamma_i$ is $p_i=\tau_i/\sum_i \tau_i$ so that
 \be
 \lambda_{0}=<\gamma_i>= {\sum_i \tau_i \gamma_i \over \sum \tau_i }
  = \, {1 \over {\avd{\tau}} }  \,  \ln
  \left( { \Delta \over \delta_0} \right) \ ,
  \qquad      \avd{\tau}=\lim_{N \to \infty} {1 \over N} \sum_{i=1}^N \tau_i
 \ee
 This definition without any modification can
 be extended to noisy systems
 by introducing the rate

 \be
       K_{\sigma}= \, {1 \over {\avd {\tau}} }  \,  \ln
  \left( { \Delta \over \delta_0} \right)
 \label{7}
 \ee
 which coincides with $\lambda_{0}$  for a deterministic system
  ($\sigma=0$). When $\sigma=0$
  there is no reason to determine the Lyapunov exponent
 in this apparently odd way,  of course.
 However,  the introduction of $K_{\s}$ is rather natural
 in the framework of the information theory \cite{a14}.
 Considering again the noiseless situation, if one wants
 to transmit the sequence $x(t)$  ($t=1,2,...T_{max}$)
 accepting only errors smaller than a tolerance threshold
 $\Delta$, one can use the following strategy:

 \begin{enumerate}
 \item[(1)] Transmit the rules which specify
 the dynamical system (1), using a finite number of bits which does not
 depend on the length $T_{max}$.
 \item[(2)] Specify the initial condition
 with precision $\delta_0$ using a number of bits
  $n=\ln_2(\Delta/\delta_0)$ which permits to arrive up to the time $\tau_1$
  where the error equals $\Delta$.
 Then specify again the new initial condition $x(\tau_1+1)$ with a
 precision $\delta_0$ and so on. The number of bits necessary to specify
 the sequence with a tolerance $\Delta$ up to $T_{max}=\sum_{i=1}^N \tau_i$ is
 $\simeq N n$ and the mean information for time step is
 $\simeq N n/T_{max}=K_{\s}/\ln 2$  bits.
 \end{enumerate}

 In presence of noise,
 the strategy of the transmission is unchanged but since
 it is  not possible to transmit the realization of
 the noise $w(t)$, one has to estimate the growth of the error
  $\delta x(t)=x'(t)-x(t)$,
 where $x(t)$ and $x'(t)$ evolve in two different noise realizations
   $w(t)$ and $w'(t)$, and $|\delta x(0)|=\delta_0$.

 The resulting equation for the evolution of $\delta x(t)$ is:
 \be
 \delta x(t+1)\simeq f'[x(t),t] \, \delta x(t) + \sigma {\tilde w}(t)
 \qquad  {\tilde w}(t)=w'(t)-w(t)
 \label{8}
 \ee

 For sake of simplicity we discuss the case
  $|f'[x(i),i]|=const=\exp \lambda_{0}$, where  (\ref{8})
  gives the bound on the error:
 \be
 |\delta x(t)| \, <
  \, e^{\lambda_0 t} \,  (\delta_0+ { \tilde {\sigma} })
 \qquad \,\, {\rm with} \
 {\tilde {\sigma}}={ {2 \sigma} \over e^{\lambda_{0}}-1}
 \label{9}
 \ee
 This formula shows that $\delta_0$ and $\tau=\avd{\tau}$  are
  not independent variables but they are  linked by the relation
 \be
 e^{\lambda_0 \tau } (\delta_0 + {\tilde {\sigma}}) \simeq \Delta
 \ee
 As a consequence, we have only one free parameter, say $\tau$,
  to optimize the information entropy $K_{\s}$ in (\ref{7}),
   so that the complexity of the
  noisy system can be estimated by
 \be
 G_{\s}=\min_{\tau} \, K_{\s} \, = \,
   \lambda_{0} +O(\sigma/\Delta)
 \ee
 where the  minimum is reached  at an optimal time
  $\tau=\tau_{opt}$ from the transmitter point of view.

 In the case of a deterministic system $K_{\s}$ does not depend
 on the value of ${\tau}$ (i.e. it is equivalent to use a long
 ${\avd \tau}$ and to transmit many bits few times or a short $\avd{ \tau}$
   and to transmit few bits many times).
 On the contrary, in noisy systems there exists an optimal time
  $\tau_{opt}$ which  minimizes $K_{\s}$:  using
 relation (\ref{9}) one sees that
   $\Delta=\exp (\lambda_{0} {\avd \tau})  ( \delta_0+ { \tilde {\sigma} })$
  and $K_{\s}$ has a minimum for
  $\tau_{opt}\simeq 1/\lambda_{\sigma}$.
 This result might appear trivial but  has
 a relevant consequence from a theoretical point of view
 in presence of noise, even if the value of the entropy
  $G_{\s}$ changes only $O(\sigma/\Delta)$, there exists an optimal time
 for the transmission.

 The interesting situation happens
  for strong intermittency when there is
  an alternation of positive and negative $\gamma$
 during  long time intervals.
  In this case the existence of an optimal time for the transmission
 induces a dramatic change for the value of $G_{\s}$.
 This becomes particularly clear when  considering the limit case
  of positive $\gamma_1$ in an interval $T_1>> 1/\gamma_1$
 followed by a negative
 $\gamma_2$ in an interval $T_2>>1/|\gamma_2|$,
 and  again a positive effective Lyapunov
 exponent and so on.
  In the expanding  intervals,  one can transmit the sequence
 using $\simeq T_1/(\gamma_1 \ln 2)$ bits, while during
  the contracting interval one
 can use only few bits.  Since  in the expanding intervals,
  the transmission has to be repeated rather often and
  moreover $|\delta x|$ cannot be lower than the noise amplitude
 $\sigma$,
  at difference with the noiseless case,
  it is impossible to use the contracting
 intervals to compensate the expanding ones.
 This implies that in the limit of very large $T_i$ only the expanding
 intervals contribute to the evolution of the error $\delta x(t)$
  and  the information entropy is given by an average
  of the positive effective Lyapunov exponents:
 \be
    G_{\s} \simeq
  <\gamma \, \theta(\gamma)>
 \label{12}
 \ee
  For the approximation considered above,
  $G_{\s} \ge \lambda_{\sigma}=<\gamma>$.
 Note that by definition
  $G_{\s} \ge 0$ while $\lambda_{\sigma}$ can be negative.
 The estimate (\ref{12}) stems from the fact that $\delta_0$
 cannot be smaller than $\sigma$ so the typical value of  $\tau_i$ is
  $O(1/\gamma_i)$ if $\gamma_i$ is positive.
  We stress again that (\ref{12}) holds
  only for strong intermittency, while
  for uniformly  expanding systems or
  rapid alternations of contracting and expanding behaviors
  $G_{\s} \simeq \lambda_{\sigma}$.

 It is not difficult to estimate the range of validity of the
 two limit cases  $G_{\s} \simeq \lambda_{\sigma}$
  and  $G_{\s} \simeq <\gamma \, \theta (\gamma)>$.
  Denoting by $\gamma_{+}>0$ and  $\gamma_{-}<0$ the typical
 values of the effective Lyapunov exponent in the expanding and contracting
 time intervals of length $T_{+}$ and $T_{-}$ respectively,
 (\ref{12}) holds if during the
 expanding intervals there are at least two repetitions of the transmission
 and  the duration of the contracting interval is  long  enough
  to allow the noise  to be dominant
  with respect to the contracting deterministic
 effects. In practice one should require

 \be
       \exp  \left(\gamma_{+} \, {T_{+} } \right)
 \gg {\Delta \over \sigma}
       \qquad
       \exp \left(-|\gamma_{-}| \, {T_{-} } \right)
 \gg {\Delta \over \sigma}
 \label{exp}
 \ee

 In a similar way,  $K\simeq \lambda_{0}$ holds
 if:
 \be
       \exp \left(-|\gamma_{-}| \, {T_{-} } \right)
 \ll {\Delta \over \sigma}
 \label{exp2}
 \ee
 We report the results of some numerical simulations
  in two  different systems which are
  showed in fig. 2, and 3, respectively.
 Let us stress that we have directly computed $K_{\s}$, and since
 $\tau_i=O(1/\gamma_i)$, we automatically are very close to the
 optimal strategy so  that $K_{\s} \simeq G_{\s}$,
 without performing a minimization.
  The random perturbation $w(t)$
  is an independent variable uniformly distributed
  in the interval $[-1/2 \, , \, 1/2]$.

  The first system is given by periodic alternation of
  two piecewise linear maps of the interval $[0,1]$
  into itself:
 \be
     f[x,t]=\cases {
 a \, x \qquad  {\rm mod}  \, 1 \qquad {\rm if} \
  (2n-1)T\le t < 2nT ; \cr
  b \, x \qquad  \qquad   \qquad {\rm if} \
   2nT \le t < (2n+1)T
 \cr}
 \label{periodic}
 \ee
 where $a>1$ and $b<1$.
  Note that in the limit of small $T$,
 $G_{\s} \to \max [\lambda_{\s}, 0]$
  since it is a non-negative quantity
  as shown in fig.(2) where for $b=1/4$ ,
  $\lambda_{\s}$ is negative.

 The second system is strongly intermittent
  without an external forcing. It is  the
 Beluzov-Zhabotinsky map [4,7]  related to a famous chemical reaction:

 \be
 f(x)= \left\{
 \begin{array}{l}
       \left[ (1/8-x)^{1/3}+a \right] e^{-x}+b \qquad {\rm if} \
         0 \le x < 1/8 ; \\
        \left[ (x-1/8)^{1/3}+a \right] e^{-x}+b \qquad {\rm if} \
        1/8 \le x < 3/10 ; \\
        c(10\, x \, e^{-10 x/3})^{19} +b \qquad {\rm if} \
        3/10 \le x
 \end{array}
 \right.
 \label{beluzov}
 \ee
 with $a=0.50607357, b=0.0232885279, c=0.121205692$.
 The map exhibits a chaotic alternation of expanding and very
 contracting time intervals.
 Although the value of $T_{-} $ is very small because $|\gamma_{-}|>>1$,
   the first inequality (\ref{exp}) is unsatisfied because
  the expanding time interval are rather short.
  As a consequence the asymptotic estimate
  $G_{\s} \simeq <\gamma \, \theta(\gamma)>$ cannot be observed.
  In fig 3, one sees that while $\lambda$ passes from
  negative to positive values at decreasing $\s$,
   $G_{\s}$ is no sensitive to this transition
  to `order'.
 Another important remark is that
 in the usual treatment of the experimental data,
 if some noise is present, one practically computes
  $G_{\s}$ and the result can be completely different from
 $\lambda_{\sigma}$. Let us mention for example \cite{a6} where the
 author studies a one-dimensional nonlinear time-dependent Langevin
 equation. A numerical computation shows that $\lambda_{\sigma}$
 is negative while the author claims to find, using the Wolf method,
  a positive `Lyapunov exponent'.

 Our results show that the same system can be regarded
 either as regular (i.e. $\lambda_{\sigma}<0$) when the
 same noise realization is considered for two nearby trajectories
 or as chaotic (i.e. $G_{\s}>0$) when two different
 noise realizations are considered.
 The situation is similar to what observed in fluids with
 lagrangian chaos.
 There, a pair of  particles passively advected by
  a chaotic  velocity field might  remain closed following
  together a `complex' trajectory. The
  lagrangian  Lyapunov exponent is thus zero. However
  a data analysis gives a positive Lyapunov exponent
 because of the `eulerian' chaos.
 We can say that $\lambda_{\sigma}$ and $G_{\s}$ correspond
 to the lagrangian Lyapunov exponent and to the exponential rate
 of separation of a particle pair  in two slightly different
 velocity fields,  respectively.

  The relation $G_{\s} \simeq  <\gamma \, \theta(\gamma)>$
  is, in some sense, the time analogous of the Pesin
  relation $h \simeq \sum_i \lambda_i \, \theta(\lambda_i)$
  between  the Kolmogorov-Sinai entropy $h$
  and the Lyapunov spectrum \cite{a16} where
  the negative Lyapunov exponents do not decrease the value of
 $h$ since the contraction along the corresponding directions
  cannot be observed for any finite space partition.
  In the same way the contracting time intervals, if
  long enough, do not decrease $G_{\s}$.

 It is important to note that the limit $\sigma \to 0$ is
  very delicate.
 Indeed for small $\s$, say $\s < \s_c$ , the inequality (\ref{exp2})
 will be fulfilled and $G_{\s} \simeq \lambda_{\s}
 \to \lambda_0$ for $\s \to 0$. However in strongly intermittent systems
 ${T}_{-}$ can be very long so that the noiseless
 limit $G_{\s} \to \lambda_0$
 is practically unreachable, as illustrated by fig 3.

 \newpage

 \section{Complexity in random dynamical systems}

 In this section we discuss dynamical systems
 (mainly maps) where the randomness is not simply given
 by an additive noise, as in sect.3.
 This kind of systems has been the subject of much interest
 in the last few years in relation with problems involving
 disorder \cite{broeck,kapral}, the characterization of
 the so-called {\em on-off intermittency} \cite{onoff}
 and the modelling of transport problems in turbulent
 flows \cite{yuott}.
 In these systems, in general, the random part represents an ensemble
 of hidden variables, that is unknown observables, believed
 to be implicated in the dynamics: the turbulent convection in
 the solar cycle or several economic factors for the stock market
 prices are just two examples of this situation.
 The random part can, also, represents the effect of a set of variables
 which vary in a chaotic way or that vary on a time scale very
 small respect to the time scale of the phenomenon under
 investigation.
 Random maps exhibit very interesting  features ranging from stable
 or quasi-stable behaviours, to chaotic behaviours and intermittency.
 In particular {\em on-off intermittency} is an aperiodic switching between
 static, or laminar, behaviour and chaotic bursts of oscillation. It can be
 generated by systems having an unstable invariant manifold, within
 which is possible to find a suitable attractor (i.e. a fixed point).
 The intermittency is linked, in the simplest case, to the
 loss of stability of the fixed point.
 For further details we refer to \cite{onoff}.

 A random map can be defined in the following way.
 Denoting with $\b{x}(t)$ the state of the system
 at the discrete time $t$, the evolution law is given by

 \begin{equation}
 {\displaystyle \bf x}(t+1)= {\displaystyle \bf F}({\displaystyle
\bf x}(t), I(t))
 \label{rap}
 \end{equation}

 where $I(t)$ is a random variable (r.v.).
 If the r.v. $I(t)$ is discrete with an entropy $h_s$,
 according the general ideas discussed in sect.3,
 a measure of the complexity of the evolution ca be defined
 in terms of mean number of bits that must be specified, at each iteration,
 in order to have a certain tolerance $\Delta$ on the
 knowledge of the state ${\displaystyle {\bf x}}$.

 Of course, it is possible to introduce a Lyapunov exponent
 $\lambda_{I}$, which is the analogue of $\lambda_{\sigma}$,
 computed considering the evolution of the tangent vector
 of eq.(\ref{rap}) once given the realization
 $I(1),I(2),...,I(t)$ of the random process.

 Therefore, there are two different contributions
 to the complexity:

 \begin{enumerate}
 \item[(a)] one has to specify the sequence $I(1),I(2),...,I(t)$
 which implies $h_s/ln 2$ bits per iteration;
 \item[(b)] if $\lambda_I$ is positive, one has to specify the initial
 condition $\b{x}(0)$ with a precision $\Delta \exp^{-\lambda_I T}$
 where $T$ is the time length of the evolution; it is necessary
 to give $\lambda_I\ln 2$ bits per iteration;
 if $\lambda_I$ is negative the initial condition can be specified using
 a number of bits which does not depend on $T$.
 \end{enumerate}

 Therefore, the complexity of the dynamics can be measured as

 \be
 \tilde K= h_s +\lambda_{I} \theta(\lambda_{I}),
 \label{compl}
 \ee
 where $\theta$ is the Heaviside step function.

 We stress again that a negative value of $\lambda_{I}$ does {\em not}
 implies predictability.

 \subsection{Two examples of random maps}

 As specific example, we  discuss a random map
 that is obtained from the deterministic chaotic evolution
 of a model made of two sliding blocks \cite{Huang}
 (see also \cite{narko}) on a rough surface.
 Such a model provides a good description of the dynamics
 of two coupled large segments of a fault.

 The equations of motion for the position of the two blocks
 during a slip can be written as

 \be
 \begin{array}{l}
 \ddot Y_{1}+Y_{1}+\alpha(Y_{1}-Y_{2})=
 {1 \over 1+\gamma\vert\dot Y_{1}-\nu\vert} \\
 \ddot Y_{2}+Y_{2}+\alpha(Y_{2}-Y_{1})={ \beta \over 1+\gamma\vert\dot
 Y_{2}-\nu
 \vert}
 \end{array}
 \label{terremoti}
 \ee
 while when one of the two blocks sticks,  one has
 \be
 \ddot Y_{1}=0  \qquad  \dot Y_{1}=\nu
 \ee
 or
 \be
 \ddot Y_{2}=0 \qquad \dot Y_{2}=\nu
 \ee
  respectively  if $\vert Y_{1}+\alpha\,(Y_{1}-Y_{2})\vert \, < \,1$, or
 $\vert Y_{2}+\alpha\,(Y_{2}-Y_{1})\vert \, < \,\beta$,
 where the $Y_i$ are the rescaled  displacements from the equilibrium
 position,
 $\alpha$ and $\gamma$ are related to the coupling constants and the
 friction dynamical coefficient.
 The quantity $T_{\nu}=\nu^{-1}$  is the natural (adimensional)
 time unit of the system. For details on the model see
 Ref.\cite{lacorata}.

 Although there is no randomness in the starting model one can
 obtain a random map in a new set of physically relevant variables.
 In sliding blocks models, the seismic moment (proportional
  to the released energy) is the sum of the  sliding runs during a
 single seismic event, that is

 \be
 M_{n} \, = \, \sum_{i=1}^{2} \, \vert Y_{i}(n+1)-Y_{i}(n)\vert
 \ee
 where $Y_i(n)$ is the position of the $i^{th}$ block before the
  $n^{th}$ slip.
 As shown in fig 4, the map $M_{n+1}$ versus $M_n$ of the seismic
  moment computed at subsequent events is multi-valued on the definition
domain.
  This is a general feature which must be taken into account when analyzing
  realistic signals generated from dynamical systems exhibiting
  low-dimensional chaos.

  Since some points have more than one image, an appropriate
  description of the dynamics is through a random map
  where a weight is assigned to each possible option.
 A good approximation of the deterministic evolution is obtained
 even considering the same weights for the two options.

 Another interesting example of a system which can treated in the
 framework of random maps is represented by the so-called Sandpile models
 \cite{bak2}.
 These models represent an interesting example of Self-Organized Criticality
 (SOC) \cite{bak1,dla,dbm,bs,ip}.
 This term refers to the tendency of large dynamical systems to evolve
 {\em spontaneously} toward a critical state characterized by spatial and
 temporal self-similarity.
 The original Sandpile Models are cellular automata inspired to
 the dynamics of avalanches in a pile of sand.
 Dropping sand slowly, grain by grain on a limited base,
 one reaches a situation where the pile is critical, i.e. it has a critical
 slope. That means that a further addition of sand will produce sliding
 of sand (avalanches) that can be small or cover the entire size of the
 system. In this case the critical state is characterized by scale-invariant
 distributions for the size and the lifetime and it is reached without the
 fine tuning of any critical parameter.

 We will refer in particular to the so-called
 Zhang model \cite{zha1}, a continuous version of the original sandpile
 model (the BTW model) \cite{bak2}, defined on a $d$-dimensional lattice.
 The variable on each site $E_{i}$ (interpretable as energy, sand, heat,
 mechanical stress etc.) can vary continuously in the range
 $\left[ 0,1 \right]$ with the threshold fixed to $E_{c}=1$.
 The dynamics is the following:

 \begin{enumerate}
 \item[{\bf (a)}]we choose a site in random way and we add to this site
 an energy $\delta$ (rational or irrational);
 \item[{\bf (b)}]if at a certain time $t$ a site, say $i$, exceeds the
 threshold $E_{c}$ a relaxation process is triggered defined as:

 \be
 \left\{
 \begin{array}{l}
 E_{i} \rightarrow 0 \\
 E_{i+nn} \rightarrow E_{i+nn} + \frac{E_{i}}{2d}
 \end{array}
 \right.
 \ee
 where $nn$ indicates the $2d$ nearest neighbours of the site $i$;
 \item[{\bf (c)}]we repeat point (b) until all the sites are relaxed;
 \item[{\bf (d)}]we go back to point (a).
 \end{enumerate}

 We can also define a deterministic  version of this model in which, at
 each addition time, we increase the variable of every site of a quantity
 $\delta$ and then follow the same rules as above updating all the sites
 over threshold in a parallel way.

 The dynamics of this model can be seen
 as described by a Piecewise linear Map \cite{cl}. In fact, indicating
 with $x \equiv \left\{ x_{i} \right\}_{i\in D}$ the configuration of
 the system at a certain time, where $D \subset Z^{d}$ is the bounded
 domain whose cardinality is $\mid D \mid =N^{d}$ with $N$ being the
 linear dimension of the lattice, the operator $\Delta_i$
 corresponding to a toppling at site $i$ is given by

 \be
 (\Delta_i \cdot x)_{j}=x_{j}-\delta_{i,j}x_i+
 {{1}\over{2d}}
 \sum^*_{i^\prime}\delta_{i',j}x_i
 \label{delta}
 \ee
 where $\sum^*$ means the sum over the nearest neighbours site of $i$.

 Eq.(\ref{delta}) shows that the single toppling is a linear operator and
 acts as a local laplacian. The evolution of a configuration up to the time
 $t$ can be written as \cite{cl}:

 \be
 x(t)= T^{t} x = L_{x,t} x_0+\delta \sum^{t}_{s=1}L_{x,t-s+1} 1_{k(s)};
 \label{xt}
 \ee
 where $L_{x,t}$ is a linear operator defined as a suitable product of linear
 operator $\Delta$, $x_{0}$ is the initial configuration and $1_{i}$ is a
 vector in $R^{D}$ whose component $i$ is $1$ and all the others are $0$.
 $k(s)$ defines the sequence of site over which there will be the random
 addition of energy at the time $s$.
 The (\ref{xt}) shows as the evolution of the Zhang model can be seen
 as the sequential application of maps, chosen, time by time, in a random
 way. Sandpile models, thus, belong to the wide class of the random maps.

 \subsection{A toy model: one dimensional random maps}

 Let us discuss a random map  which, in spite of its
 simplicity, captures some basic features of this kind of systems:

 \be
 x(t+1)=a_t x(t) (1-x(t))
 \label{rap2}
 \ee
 where $a_t$ is a random dichotomic variable given by

 \be
 a_t= \left\{
 \begin{array}{l}
 4 \phantom{aaaaa} \mbox{with probability $p$}\\
 1/2 \phantom{aaa} \mbox{with probability  $1-p$}
 \end{array} \right.
 \label{rap3}
 \ee

 \noindent
 It is easy to understand the behaviour for $x(t)$ close to zero.
 The solution of (\ref{rap2}), keeping the linear part is:

 \be
 x(t)= \prod_{j=0}^{t-1} a_j x(0).
 \ee

 The long-time behaviour of $x(t)$ is given by the product
 $\prod_{j=0}^{t-1} a_j$.
 Using the law of large numbers one has that the typical behaviour is

 \be
 x(t) \sim x(0) e^{<ln a> t}.
 \ee

 Since $<ln a> =p ln 4+(1-p) ln 1/2=(3p-1) ln 2$
 one has that, for $p < p_{c} =1/3$,  $x(t) \rightarrow 0$ for
 $t \rightarrow \infty$. On the contrary for $p > p_c$ after a certain
 time $x(t)$ is far from the fixed point zero and the non-linear
 terms are relevant.
 Fig.(5) shows a typical {\em on-off intermittency} behaviour
 for $p$ slightly larger than $p_c$.

 Let us note that, in spite of this irregular behaviour, numerical
 computations show that the Lyapunov exponent $\lambda_I$ is negative for
 $p < \tilde p_c \simeq 0.5$: this is essentially due to the non-linear
 terms.

 For such a system with {\em on-off intermittency} it is possible,
 in practice, to define a complexity of the sequence which turns out to be
 much smaller than the value given by the general formula (\ref{compl}).

 Let us denote with $l_L$ and $l_I$ the average life times respectively of
 the laminar and of the intermittent phases for $p$ close to $p_c$
 ($l_I << l_L$).
 It is easy to realize that the mean number of bits,
 per iteration, one has to specify in order to transmit the sequence is:

 \be
 \tilde K \simeq \frac{l_I h_s}{(l_I+l_L) ln 2}
 \simeq \frac{l_I}{l_L}  \frac{h_s}{ln2}.
 \label{compl2}
 \ee

 \noindent
 The previous formula is obtained noting that on an interval $T$ one has
 $\simeq \frac{T}{l_I+l_L}$ intermittent bursts.
 Since during the intermittent bursts, i.e. $x(t)$ far from zero,
 there is not an exponential growth of the distance between two
 trajectories initially close and computed with the same sequence of $a_t$.
 So, one has just to give the sequence of $a_t$ on the intermittent bursts.
 Eq.(\ref{compl2}) has an intuitive interpretation: in systems with
 a sort of "catastrophic"  events, the most important feature is the
 mean time between two subsequent events.

 \newpage

 \section{The case of Sand-piles models}

 In this section we discuss the problem of the predictability in
 Sandpile Models \cite{bak2},

 Different authors \cite{} suggested that Self-Organized Critical systems
 occupy a particular position among the dynamical systems which has
 been named Weak Chaos. This because it has been argued that the
 maximum Lyapunov exponent of these systems is zero.
 From this it is deduced that two initially close trajectory in the phase
 space will diverge just algebraically in time and not in an exponential way,
 as do the chaotic systems. From this point of view these systems
 would seem more predictable than chaotic systems in that a better knowledge
 of the initial conditions would considerably improve the predictability time
 $T_p$

 \be
 T_p \simeq (\Delta_{max} / \delta_0)^{\alpha}.
 \label{tpalg}
 \ee

 where $\delta_0$ is the error on the
 determination  of the initial conditions, $\Delta_{max}$ is the maximum
 tolerance between the real evolution and the simulation that makes any
 prediction and $\alpha$ is just the exponent of the algebraic divergency
 of the error.
 We recall that for chaotic systems the predictability time is given by

 \be
 T_p=1/\tau \cdot ln (\frac{\Delta_{max}}{\delta_0}).
 \label{tpesp}
 \ee

 In this case an improvement in $\delta_0$ would increase $T_p$ just in
 a logarithmic way.

 In this context we would like to discuss this problem on the basis of some
 recent rigorous results \cite{cl} in order
 to clarify the role of the Lyapunov exponents for these class of systems
 and to address the problem of the predictability.

 We will refer to the Zhang model defined in sect. 4.1.

 The evolution of a configuration up to the time $t$ is given by the
 (\ref{xt}) which shows as the evolution of the Zhang model can be seen
 as the sequential application of maps, chosen, time by time,
 in a random way.

 The Lyapunov exponent corresponding to a
 given trajectory $x(t)= T^{t} x$ can be defined, linearizing the dynamics
 in the neighborhood of $x(t)$, as (\cite{bg}):

 \be
 \lambda
 \equiv
 \lim_{t\rightarrow\infty}
 {{1}\over{t}}ln{{\vert z(t) \vert}
 \over{\vert z(0) \vert}};
 \label{lamdef2}
 \ee

 where $z(t)$ represents the distance between two different configurations
 $x$ and $y$ at the time $t$.
 In example, in the $L^1$ norm $z(t)=\sum_i \vert y_i(t)-x_i(t) \vert$
 with $i=1,N^d$.
 If the two trajectories $x(t)$ and $y(t)$
 make the same sequence of toppling eq.(\ref{lamdef2}) holds with the
 substitution $y-x \rightarrow z$. In fact, in this case, it holds
 $T^{t}y -T^{t}x= T^{t}z  =z(t)$. Therefore the definition
 (\ref{lamdef2}) for the Lyapunov exponent fail when the
 two configuration begin to follow different sequences of toppling.
 It is easy to see that such a  situation occurs when, for one
 configuration, it holds $x_{i}(t)=1$ for some $i$ and $t$.
 In this case a little difference in the second configuration
 $y_{i}(t)=x_{i}(t)+\epsilon$ will produce a toppling just in
 the $y$ configuration. From this point onwards the two configurations
 will follow different sequences and the definition (\ref{lamdef2})
 fails definitely.

 It is easy to see that the Lyapunov exponent is not positive.
 In fact, the dynamics in the tangent space, the dynamics of a
 little difference between two configurations, follows the same
 rules of the usual dynamics and the "error" is redistributed to
 the nearest neighbours site.

 It is then clear that the distance between two configurations,
 being conserved in the toppling far from the boundaries, can
 just decrease when a site of the boundary topples. We can conclude
 that $\lambda \le 0$.

 In \cite{cl} it has been obtained rigorously that, for
 the maximum Lyapunov exponent $\lambda$, as defined in (\ref{lamdef2}),
 it holds:

 \be
 \lambda \le - \frac{1}{ N^d (R(D)+1)^{2}(1/ \delta +1)
 (log N^d +1)}
 \ee

 where we indicated with $R(D)$ the diameter of the domain $D$,
 that is that the Lyapunov exponent is strictly
 lower than $0$.

 An immediate consequence of this Theorem is that
 the dynamics, up to the time $t$ ( for $t$ sufficiently large) is
 given by a {\em Piecewise Linear Contractive Map}.

 At first, one could think that the existence of
 a negative Lyapunov exponent should assure a perfect predictability.
 That is not true. What makes the situation complex is the existence of
 a splitting mechanism in the configuration space which affect the so-called
 snapshot attractor.
 A snapshot attractor is obtained by considering a cloud of
 initial conditions and letting it evolve forward in time
 under a given realization of the noisy dynamics.
 We can identify two different mechanism which concur to the
 formation of the snapshot attractor:

 \begin{enumerate}
 \item[{\bf (a)}] a volume contraction mechanism due to the effect of
 the negative Lyapunov exponent;
 \item[{\bf (b)}] a splitting mechanism which tends, by virtue of the
 piecewise structure of the map, to map single sets of configurations
 in two or more distinct sets also far apart in the phase space.
 \end{enumerate}

 The splitting mechanism (b) tends to create a partition of the
 configuration space in regions which follow the same sequence of
 toppling, whereas mechanism (a) tends to contract the volumes of
 the elements of the partition.

 It is worth to stress how, in same cases,
 it happens that the evolution
 of all the possible configurations shrink to the evolution of a single
 configuration (a point in the configuration space) whose evolution
 corresponds, at each time, to a snapshot attractor given by just one point.

 This happens, in example, in the case of a on-dimensional (linear) chain of
 $L$ sites driven with an arbitrary $\delta$. In \cite{cl} it has
 been studied the case in which $\delta=1/2$ and it has been shown that
 in this case the partition is time-independent.
 Let us discuss, for sake of simplicity and without loss of generality,
 this last case.
 A certain cloud of configurations, i.e. belonging to a same element of
 the partition, will evolve in a cloud of configurations, in principle
 smaller than the initial one due to the contractivity of the map, belonging
 entirely to another element of the partition; at its turn this cloud
 will evolve in a smaller cloud of configurations belonging to another
 element of the partition and so on. This process continues until
 all the configurations shrank to just one that continues to evolve
 jumping between different elements of the partition and evolving
 according to the map corresponding to each element of the partition.
 The Lyapunov exponent, in this case, gives informations about
 the rate of shrinking of the different clouds of configurations,
 i.e. it gives the typical exponential contracting rate of the radius
 of the snapshot attractor.

 The rigorous study of the properties of the snapshot attractors is
 out of the purposes of present work and it will be treated elsewhere.
 Here we just want to note how this situation does not change the problem
  of the predictability in that, in order to forecast the system,
 one should be able to know the random sequence which drives it.

 This puts the problem of the definition of a predictability in a
 wider perspective in which the Lyapunov exponent is not the only
 relevant quantity. Since the Lyapunov exponent gives informations
 only at very large time and for infinitesimal perturbations
 the dynamical balance of the two effects (a) and (b)
 represents a basis for the definition of a predictability for
 such a systems.

 Let us consider initially the situation in which two different
 configurations are driven with the same realization of the noise:
 that means that at each time the sand (energy) is added to the same sites
 for the two configurations.

 Up to the time in which two different
 configurations make the same sequence of toppling the error
 $\epsilon$ ( the distance between the two configurations) will decrease.
 When the configurations begin to follow a different sequence of
 toppling the error $\epsilon$ become of order $1$ in a single time
 step whatever was $\epsilon$ before this time.
 From this point onwards the evolution of the distance between
 two configurations seems far from being linked to the Lyapunov exponent.
 The threshold mechanism, and then the splitting mechanism,
 plays therefore a crucial role in determining the predictability of
 such a systems. The system remains  definitely predictable up to
 the time in which two different configurations make the same
 sequence of toppling. This time can be defined as the predictability
 time. For a more complete treatment of this point we refer to \cite{cl2}.
 In particular it is possible show how a predictability
 for such a class of models can be related to a threshold mechanism,
 in which the Lyapunov exponent is not the only relevant quantity.
 Smaller is the initial distance between the two configurations, smaller
 will be the probability of different toppling.
 The threshold in the initial distance between the two configurations,
 say $\epsilon_T$, has, in this case, a probabilistic value.
 If $\epsilon(t=0) < \epsilon_{T}$ we cannot exclude the possibility
 that the configurations will follow different sequences of
 toppling, but the probability associated
 with this event become exponentially small as the time goes on because,
 due to the negative Lyapunov exponent, two different configurations
 tend to converge each other.

 In order to confirm these predictions we simulated the parallel evolution of
 two different configurations in the random case (for a system with $L=30$)
 with different starting error $\epsilon$ and we plotted the distance
 (in the $L^{1}$ norm) between the two orbits.
 Fig.(6-a,b) show the
 results respectively for $\epsilon = 10^{-2}$ and $10^{-3}$.
 These results seem to confirm the existence of  a probabilistic
 threshold in $\epsilon$ which determines the divergence  or the
 asymptotic convergence of two orbits.

 \subsection{Predictability with different realizations of the randomness}

 It is very interesting to investigate what happens when one considers
 the case, more relevant from the point of view of the predictability,
 in which two configurations are driven by different randomness.
 That means that at each time the sand is added in different sites
 for the two realizations. Obviously we can imagine a situation in which
 the uncertainty in the knowledge of the noise can be varied. In fact
 chosen a site for one configuration, we can drive the other configuration
 putting the sand in a site which can  be one of the nearest neighbours
 sites, or one of the second nearest neighbours sites etc. of the site
 of the first configuration. In this way, due to the
 discrete structure of the system, the uncertainty cannot be reduced at will.
 The minimum uncertainty is obtained putting the sand in one of the nearest
 neighbours of the site chosen for the first configuration.
 In our simulations we considered this last situation. The results are shown
 in fig.(7). As it is possible to see, the situation in this case is much
 more involved and the threshold mechanism, described above for the case of
 the same realizations of the randomness, does not hold anymore.
 That is because, in this case, the two configurations can start to follow
 different sequences of toppling at the first toppling. From this point
 onwards we return to the situation in which, with the same realization
 of the noise, the two configurations start to follow different sequences of
 toppling; the system become unpredictable.
 Also in this case the Lyapunov exponent does not play the role of the only
 relevant quantity.

 In order to better explain this point it is possible to define the
 complexity $K$ for such kind of systems.
 In this case we can use the (\ref{compl}) of Sect.4  which we write
 again, for sake of clarity:

 \be
 \tilde K= h_s +\lambda_{I} \theta(\lambda_{I}),
 \label{compl3}
 \ee

 where $h_s$ defines the complexity relative to the choose of
 the random sequence of addition of energy.
 In Sandpile models, for example,
 since each site has the same probability to be selected,
 one has $h_s=log L$, where $L$ is the number of sites of the
 system and the second term does not exist in that
 the Lyapunov exponent is negative.
 We then obtain the result that the complexity for sandpile models
 is just determined by the randomness in the choose of the sequence
 of addition of energy;
 nevertheless, once this sequence is known, the system could be,
 all the same, unpredictable, at least for some initial conditions
 and non inifinitesimal perturbations, due to the splitting mechanism
 cited above.
 Once more, let us emphasize that a negative Lyapunov exponent implies
 predictability only if the sequence which drives the system is exactly known.

 \newpage

 \section{Conclusions}

 In this paper we focus the problem of an appropriate definition
 of the concept of complexity in random dynamical systems.

 At first,  one could follow a naive approach where the randomness
 is considered as a standard time-dependent term. In this way,
 the Lyapunov exponent $\lambda_\sigma$ is given by the rate
 of divergence of two initially close trajectories
 evolving under the same realization of the randomness.

 Although well defined from a mathematical point of view, such an approach
 leads to paradoxical situations. For instance, in
 a system driven by the
 one-dimensional Langevin equation,
 the existence of a negative Lyapunov exponent does not imply
 the possibility to forecast the future state of the system
 unless one  exactly knows the realization of the noise.
 Another paradox is represented by the situation discussed in sect.2.2 where
 two different systems, one with positive Lyapunov exponent, and the other
 with a negative one, appear practically undistinguishable.
 Least but not last,
 it is practically impossible to extract $\lambda_\sigma$
 from an analysis of experimental data.

 The main result of the paper is the definition of
 a measure of complexity $K$ in terms of the
 mean number of bits per time unit necessary to specify the
 sequence generated by the random evolution law.
 We have also shown that from a practical point of view, this
 definition correspond to consider the divergence of nearby
 trajectories evolving in different noise realizations.
 The great advantage is
  that $K$ can be extracted from experimental data \cite{a12}.
 The two indicators $K$ and $\lambda_\sigma$
 have a close values and are practically equivalent in systems with
 weak dynamical intermittency. However, in presence of strong
 intermittency (say irregular alternations of long
 regular periods with sudden chaotic bursts)
 $K$ and $\lambda_\sigma$ become very different and in extreme situations
 it may happen that  $K$ is positive  while $\lambda_\sigma$ negative.
 It is thus questionable whether such a system is chaotic or regular
 and to speak of noise induced order.

 A special class of systems well described by
 our characterization are random maps, where
   at each time step, different possible evolution
 laws are chosen  according to a given probabilistic rule.
 Sandpile models form an important  group
 of systems tan can be described in terms of random maps.
  The existence of a negative Lyapunov exponent
 does not allow one to capture the basic features
of these spatially extended
 systems
  while our measure of complexity
 is able to describe in an appropriate way the dynamical behaviour.

It seems to us  that the study
 of the complexity and of the predictability
 is completely understood only in the case of deterministic
 dynamical systems with few degrees of freedom.
 Our work wants to be a first step toward
 a deeper comprehension of these issues
 in systems with many degrees of freedom or
  in interactions with many degrees of freedom
 represented by a noise, problems that
  are still open and sometimes controversial.

\vspace{1.0cm}

{\bf \large Acknowledgements}

It is a pleasure to thank E. Caglioti for useful suggestions and
discussions. We also give thanks to R. Kapral and T. Tel for the useful
correpondence.

 \newpage
 {\bf \large Figure captions}

 \begin{itemize}

 \item[{\bf Fig.1:}] {\bf (a)} $z(t)$ vs. $t/T$ for the system
 (\ref{season}): $R_0 =25.5, A=4.$ and $T=1600$.
 {\bf (b)} $z(t)$ vs. $t/T$ for the system (\ref{langevin}):
 $R_0 =20.0, A=5.$, $T=1600$ and $\sqrt{2 \sigma}=0.15$.

 \item[{\bf Fig.2:}]
 $K_{\sigma}$ versus $T$ with $\sigma=10^{-7}$ for the map
 (\ref{periodic}). The parameters of map (\ref{periodic})
 are $a=2$ and $b=2/3$ (squares) or $b=1/4$ diamonds.
 The dotted line indicates the Pesin-like relation (\ref{12})
 while the dashed lines are the noiseless limit of $K_{\sigma}$.
 Note that for $b=1/4$ the Lyapunov exponent $\lambda_{\sigma}$ is negative.

 \item[{\bf Fig.3}] $\lambda_{\sigma}$ (squares) and $K_{\sigma}$
 (crosses) versus $\sigma$ for map (\ref{beluzov}).

 \item[{\bf Fig.4}] Multi-valued map of the seismic moments:
 $M_{n+1}$ versus $M_n$ generated by equations (\ref{terremoti})
 where $\beta=2.0, \alpha=1.2$ and $\gamma=3.0$.

 \item[{\bf Fig.5}] $x(t)$ vs. $t$ for the random map
 (\ref{rap2}, \ref{rap3}) with $p=0.35$.

 \item[{\bf Fig.6:}] Evolution of the distance $\epsilon$
 between two configurations driven with the same realization
 of the randomness and with a starting distance of:
 {\bf (a)} $10^{-2}$ and {\bf (b)} $10^{-3}$.

 \item[{\bf Fig.7:}] Evolution of the distance $\epsilon$
 between two initially identical configurations driven with different
realizations of the randomness.

 \end{itemize}

 \end{document}